
%
%
%
\input phyzzx
\tolerance=1000
\sequentialequations
\def\rl{\rightline}

\def\r#1{$\bf#1$}

\def\t1{{\tilde 1}}

\def\AEF{A.E. Faraggi}
\def\DVN{D.V. Nanopoulos}

\def\SUSY{supersymmetry }

\def\NPB#1#2#3{Nucl. Phys. B {\bf#1} (19#2) #3}
\def\PLB#1#2#3{Phys. Lett. B {\bf#1} (19#2) #3}
\def\PRD#1#2#3{Phys. Rev. D {\bf#1} (19#2) #3}

\def\PRT#1#2#3{Phys. Rep. {\bf#1} (19#2) #3}

\def\l{\langle}
\def\r{\rangle}

\REF\SUSY{H. P. Nilles, For reviews see,
H.P. Nilles, \PRT{110}{84}{1}; \DVN~ and Lahanas, \PRT{145}{87}{1}.}
\REF\W{S. Weinberg, \PRD{26}{82}{475}.}
\REF\SAKAI{S. Sakai and T. Yanagida,
\NPB{197}{82}{533}; R. Arnowitt, A.H. Chamsdine and P. Nath,
\PRD{32}{85}{2348}.}
\REF\DRW{S. Dimopoulos, S. Raby and F. Wilczek, \PRD{24}{81}{1681}.}
\REF\G{B. Grinstein, \NPB{206}{82}{206};
       J. Ellis, J.S. Hagelin, D.V. Nanopoulos and K. Tamvakis,
          \PLB{115}{82}{380}; I. Antoniadis, J. Ellis, J.S. Hagelin and
          D.V. Nanopoulos, \PLB{194}{87}{231}.}
\REF\GSW{M. Green, J. Schwarz and E. Witten,
Superstring Theory, 2 vols., Cambridge
University Press, 1987.}
\REF\SUTHREE{B. Greene {\it{el al.}},
Phys.Lett.{\bf B180} (1986) 69;
Nucl.Phys.{\bf B278} (1986) 667;  {\bf B292} (1987) 606;
R. Arnowitt and  P. Nath, Phys.Rev.{\bf D39} (1989) 2006; {\bf D42}
(1990) 2498; Phys.Rev.Lett. {\bf 62} (1989) 222.}
\REF\DSW{M. Dine, N. Seiberg and E. Witten, Nucl.Phys.{\bf B289} (1987) 585.}
\REF\FNY{\AEF, D.V. Nanopoulos and K. Yuan, \NPB{335}{90}{347}.}
\REF\SLM{\AEF, \NPB{387}{92}{239}.}
\REF\EU{\AEF,  \PLB{278}{92}{131}.}
\REF\TOP{\AEF, \PLB{274}{92}{47}.}
\REF\GCU{\AEF, \PLB{302}{93}{202}.}
\REF\FIQ{A. Font, L.E. Iba{\~n}ez and F. Quevedo,
 Phys.Lett.{\bf B228} (1989) 79.}
\REF\NRT{\AEF, \NPB{403}{93}{101}.}
\REF\FFF{I. Antoniadis, C. Bachas, and C. Kounnas, Nucl.Phys.{\bf B289}
(1987) 87; I. Antoniadis and C. Bachas, Nucl.Phys.{\bf B298} (1988)
586; H. Kawai, D.C. Lewellen, and S.H.-H. Tye,  Nucl.Phys.{\bf B288} (1987) 1;
R. Bluhm, L. Dolan, and P. Goddard, Nucl.Phys.{\bf B309} (1988) 330.}
\REF\REVAMP{I. Antoniadis, J. Ellis, J. Hagelin, and \DVN, \PLB{231}{89}{65}.}
\REF\KLN{S. Kalara, J. Lopez and D.V. Nanopoulos, \PLB{245}{91}{421};
\NPB{353}{91}{650}.}
\REF\SO{S. Dimopoulos, S. Raby and F. Wilczek, \PLB{112}{82}{133}.}
\REF\FBV{J. Ellis, J. Lopez and D.V. Nanopoulos, \PLB{252}{90}{53},
G. Leontaris and T. Tamvakis, \PLB{260}{91}{333}.}
\REF\KLN{S. Kalara, J. Lopez and D.V. Nanopoulos, \PLB{245}{91}{421};
\NPB{353}{91}{650}.}

\singlespace
\rl{IASSNS--HEP--94/18}
\rl{\today}
\rl{T}
\normalspace
\smallskip
\titlestyle{\bf{Proton stability in superstring derived models}}
\author{Alon E. Faraggi
{\footnote*{Work supported by an SSC fellowship.
                 e--mail address: faraggi@sns.ias.edu}}}
\smallskip
\centerline {School of Natural Sciences}
\centerline {Institute for Advanced Study}
\centerline {Olden Lane}
\centerline {Princeton, NJ 08540}

\titlestyle{ABSTRACT}
I discuss the problem of proton decay,
from dimension four, five and six operators,
in superstring derived standard--like models. I classify the sectors
that produce color triplet superfields which may generate
proton decay from dimension five and six operators. I show that for
two of these sectors there exist a unique superstring
doublet--triplet splitting mechanism that projects out the color triplets
by means of the GSO projection, while leaving the
electroweak doublets in the physical spectrum. I investigate possible
proton decay due to additional color triplets in the massless spectrum
and due to nonrenormalizable terms. I show that there exist models
in which the dimension four, five and six operators,
that lead to proton decay,
are forbidden by the symmetries of the string models, to all orders
of nonrenormalizable terms.
\centerline{}

\singlespace
\vskip 0.5cm
\nopagenumbers
\pageno=0
\endpage
\normalspace
\pagenumbers

\centerline{\bf 1. Introduction}

LEP precision data provides further support to the validity of the Standard
Model at the electroweak scale and possibly to a much higher energy scale.
In the last two decades this possibility has been exploited in the
development of Grand Unified Theories and Superstring Theories.
In general, theories of unification lead to a well known prediction,
namely proton decay. Non supersymmetric unified theories are all but ruled
out due to proton decay constraints, while their supersymmetric
counterparts must accommodate some ad hoc and very stringent symmetries,
if they are to survive proton lifetime limits. Supersymmetric
models [\SUSY], in general,
admit proton decay from dimension four, five and six
operators [\W,\SAKAI].
Dimension four operators are avoided by imposing R--parity.
Dimension six operators, from gauge boson exchange, are suppressed because
the unification scale in supersymmetric models is higher than in
their nonsupersymmetric counterparts [\DRW].
However, in supersymmetric models dimension five operators from Higgsino
exchange are the most problematic. This requires that the
Higgsino color multiplets are sufficiently heavy,
of the order of $10^{16}~GeV$.
Supersymmetric GUT models must admit some doublet--triplet splitting
mechanism, which satisfies these requirements. Although, such a mechanism
has been constructed in different supersymmetric GUT models, in general,
further assumptions have to be made on the matter content and
interactions of the supersymmetric GUT models [\G]. An important point of this
paper is to show how the doublet--triplet splitting problem may be solved
in a class of superstring standard--like models.

Superstring theories [\GSW]
lead in their point field theory limit to $N=1$ space--time
supersymmetry and therefore usually suffer from the same problems.
In the context of superstring theory the problem is worsen because
one cannot impose additional symmetries at will. The desired symmetries
must be derived in the massless spectrum of specific string vacua.
In many string models the required symmetries are obtained at
specific points in their moduli space [\SUTHREE].
However, to produce realistic low
energy mass spectrum it is in general necessary to perturb away
from the symmetric points in moduli space. In this case the operators
that produce proton decay are generated from nonrenormalizable terms.
The nonrenormalizable terms of order $N$ produce effective terms,
$\lambda ffb\phi^L\l\phi\r^{N-L}/M^{N-3}$, where $\l\phi\r$ are the VEVs that
break the symmetries which forbid the cubic level Baryon violating operators.
These VEVs are necessary if we impose a supersymmetric vacuum at the Planck
scale [\DSW].
Thus, although the dangerous operators are forbidden at the cubic
level of the superpotential they are in general generated from
nonrenormalizable terms. To satisfy proton lifetime limits there
must exist a mechanism that suppresses the dangerous operators,
which is generation independent and which is independent of the
particular point in moduli space.

In this paper I study the problem of proton decay in a class of
superstring standard--like models [\FNY--\TOP] that are constructed in the
free fermionic formulation [\FFF]. I argue that in string models dimension
four, baryon and lepton violating, operators are in general generated.
The existence of an additional, generation independent,
$U(1)$ symmetry which remains  unbroken down to some energy scale
is sufficient to solve this problem [\W,\FIQ,\NRT].
I show that in the superstring
standard--like models there is a unique doublet--triplet
splitting mechanism. The dangerous color triplets are projected out
from the physical spectrum by the GSO projection,
while the electroweak doublets remain in the
physical spectrum.
The GSO projections are imposed
by requiring modular invariance.
The models contain additional
color triplets from sectors that are generated by the Wilson line
breaking. I show in one specific model that the symmetries of the
string model forbid the dangerous dimension five and six operators,
from Higgs and Higgsino exchange and
to all orders of nonrenormalizable terms.

\bigskip
\centerline{\bf 2. The superstring models }

The superstring standard--like models are derived in the free fermionic
formulation [\FFF].  In this formulation all the degrees
of freedom needed to cancel
the conformal anomaly are represented
in terms of internal free fermions propagating
on the string world--sheet.
Under parallel transport around a non-contractible loop,
the fermionic states pick up a phase.
Specification of the phases for all world--sheet fermions
around all noncontractible loops contributes
to the spin structure of the model.
The possible spin structures are constrained
by string consistency requirements
(e.g. modular invariance).
A model is constructed by choosing a set of boundary condition vectors,
which satisfies
the modular invariance constraints.
The basis vectors, $b_k$, span a finite
additive group $\Xi=\sum_k{{n_k}{b_k}}$
where $n_k=0,\cdots,{{N_{z_k}}-1}$.
The physical massless states in the Hilbert space of a given sector
$\alpha\in{\Xi}$, are obtained by acting on the vacuum with
bosonic and fermionic operators and by
applying the generalized GSO projections.

The superstring standard--like models are generated by a basis of
eight vectors of boundary conditions
for all the world--sheet fermions. The first five vectors in the
basis consist of the NAHE set, $\{{{\bf 1},S,b_1,b_2,b_3}\}$
[\REVAMP,\SLM]. The gauge group after the NAHE set
is $SO(10)\times SO(6)^3\times E_8$ with $N=1$ space--time supersymmetry.
The vector $S$ is the supersymmetry generator and the superpartners of
the states from a given sector $\alpha$ are obtained from the sector
$S+\alpha$. The vectors $b_1$, $b_2$ and $b_3$ correspond to the three
twisted sectors in the corresponding orbifold formulation.
In addition to the first five vectors the basis contains three additional
vectors that correspond to Wilson line in the orbifold language.
The additional vectors distinguish between different models and
determine their low energy properties.

The NAHE set divides the 44 right--moving and 20 left--moving real internal
fermions in the following way:
${\bar\psi}^{1,\cdots,5}$ are complex and produce the observable $SO(10)$
symmetry;
${\bar\phi}^{1,\cdots,8}$ are complex and produce the hidden $E_8$ gauge
group;
$\{{\bar\eta}^1,{\bar y}^{3,\cdots,6}\}$, $\{{\bar\eta}^2,{\bar y}^{1,2}
,{\bar\omega}^{5,6}\}$, $\{{\bar\eta}^3,{\bar\omega}^{1,\cdots,4}\}$
 give rise to the three horizontal $SO(6)$ symmetries.
The left--moving $\{y,\omega\}$ states
are divided to,
$\{{y}^{3,\cdots,6}\}$, $\{{y}^{1,2}
,{\omega}^{5,6}\}$, $\{{\omega}^{1,\cdots,4}\}$.
The left--moving $\chi^{12},\chi^{34},\chi^{56}$ states
carry the supersymmetry charges.
Each sector $b_1$, $b_2$ and $b_3$ carries periodic boundary conditions
under $(\psi^\mu\vert{\bar\psi}^{1,\cdots,5})$ and one of the three groups:
$(\chi_{12},\{y^{3,\cdots,6}\vert{\bar y}^{3,\cdots6}\},{\bar\eta}^1)$,
$(\chi_{34},\{y^{1,2},\omega^{5,6}\vert{\bar y}^{1,2}{\bar\omega}^{5,6}\},
{\bar\eta}^2)$ and
$(\chi_{56},\{\omega^{1,\cdots,4}\vert
{\bar\omega}^{1,\cdots4}\},{\bar\eta}^3)$.
The division of the internal fermions is a reflection of the underlying
$Z_2\times Z_2$ orbifold compactification. The set of
internal fermions ${\{y,\omega\vert{\bar y},{\bar\omega}\}^{1,\cdots,6}}$
corresponds to the left--right symmetric conformal field theory of the
heterotic string, or to the six dimensional compactified manifold in a
bosonic formulation. This set of left--right symmetric
internal fermions plays a fundamental role in the determination of the
low energy properties of the superstring standard--like models. In
particular, it plays an important role in the superstring
doublet--triplet splitting mechanism.

The observable gauge group after application of the generalized GSO
projections is
$SU(3)_C\times U(1)_C\times SU(2)_L\times U(1)_L
\times U(1)^3\times U(1)^n${\footnote*{
$U(1)_C={3\over 2}U(1)_{B-L}$ and
$U(1)_L=2U(1)_{T_{3_R}}$.}}. The weak hypercharge is given by
$U(1)_Y={1\over 3} U(1)_C + {1\over 2} U(1)_L$ and the orthogonal
combination is given by $U(1)_{Z^\prime}= U(1)_C - U(1)_L$. The first
three additional $U(1)$ symmetries arise from the world--sheet complex
fermions,
${\bar\eta}^j$, $(j=1,2,3)$. The additional $U(1)^n$ symmetries
arise from complexifying two right--moving real fermions from the set
$\{{\bar y},{\bar\omega}\}^{1,\cdots,6}$.
For each right--moving gauged $U(1)$ symmetry there is a corresponding
left--moving global $U(1)$ symmetry. Alternatively, a left--moving
real fermion can be paired with a right--moving real fermion to form
an Ising model operator [\KLN].
The hidden
gauge group after application of the generalized GSO projections is
$SU(5)_H\times SU(3)_H\times U(1)^2$.
The $U(1)$ symmetries in the
hidden sector, $U(1)_7$ and $U(1)_8$,
correspond to the world--sheet currents
${\bar\phi}^1{\bar\phi}^{1^*}-{\bar\phi}^8{\bar\phi}^{8^*}$ and
$-2{\bar\phi}^j{\bar\phi}^{j^*}+{\bar\phi}^1{\bar\phi}^{1^*}
+4{\bar\phi}^2{\bar\phi}^{2^*}+{\bar\phi}^8{\bar\phi}^{8^*}$ respectively,
where summation on $j=5,\cdots,7$ is implied.

The massless spectrum of the
standard--like models contain three chiral generations
from the sectors $b_1$, $b_2$ and $b_3$ with charges under the
horizontal symmetries. Three generations from the sectors
$b_1$, $b_2$ and $b_3$ are common to all the free fermionic standard--like
models. For example in the model of Ref. [\GCU] we have,
$$\eqalignno{&({e_L^c}+{u_L^c})_{{1\over2},0,0,{1\over2},0,0}+
({d_L^c}+{N_L^c})_{{1\over2},0,0,{{1\over2}},0,0}+
(L)_{{1\over2},0,0,-{1\over2},0,0}+(Q)_{{1\over2},0,0,-{1\over2},0,0},~&(1a)\cr
&({e_L^c}+{u_L^c})_{0,{1\over2},0,0,{1\over2},0}+
({N_L^c}+{d_L^c})_{0,{1\over2},0,0,{1\over2},0}+
(L)_{0,{1\over2},0,0,-{1\over2},0}+
(Q)_{0,{1\over2},0,0,-{1\over2},0},~&(1b)\cr
&({e_L^c}+{u_L^c})_{0,0,{1\over2},0,0,{1\over2}}+
({N_L^c}+{d_L^c})_{0,0,{1\over2},0,0,{1\over2}}+
(L)_{0,0,{1\over2},0,0,-{1\over2}}+
(Q)_{0,0,{1\over2},0,0,-{1\over2}}.~&(1c)\cr}$$
 where
$$\eqalignno{{e_L^c}&\equiv [(1,{3\over2});(1,1)];{\hskip .6cm}
{u_L^c}\equiv [({\bar 3},-{1\over2});(1,-1)];{\hskip .2cm}
Q\equiv [(3,{1\over2});(2,0)]{\hskip 2cm}
&(2a,b,c)\cr
{N_L^c}&\equiv [(1,{3\over2});(1,-1)];{\hskip .2cm}
{d_L^c}\equiv [({\bar 3},-{1\over2});(1,1)];{\hskip .6cm}
L\equiv [(1,-{3\over2});(2,0)]{\hskip 2cm}
&(2d,e,f)\cr}$$
of $SU(3)_C\times U(1)_C\times SU(2)_L\times U(1)_L$
The vectors $b_1,b_2,b_3$ are the only vectors in the additive group
$\Xi$ which give rise to spinorial $16$ of $SO(10)$.

The vector $2\gamma$ breaks the $E_8$ symmetry at the level of
the NAHE set to $SO(16)$. The sectors $b_j+2\gamma$ produce $16$
representation of $SO(16)$, which, decompose under the final hidden
gauge group
after application of the $\alpha$, $\beta$ and $\gamma$ projections.
Like the sectors $b_j$, the sectors $b_j+2\gamma$ are common to all the
free fermionic standard--like models. For example in the model of Ref. [\GCU]
we have,
$$\eqalignno{&({V_1}+{T_1})_{0,{1\over2},{1\over2},{1\over2},0,0}+
({\bar V}_1+{\bar T}_1)_{0,{1\over2},{1\over2},-{1\over2},0,0},&(3a)\cr
&({V_2}+{T_2})_{{1\over2},0,{1\over2},0,{1\over2},0}+
({\bar V}_2+{\bar T}_2)_{{1\over2},0,{1\over2},0,-{1\over2},0},&(3b)\cr
&({V_3}+{T_3})_{{1\over2},{1\over2},0,0,0,{1\over2}}+
({\bar V}_3+{\bar T}_3)_{{1\over2},{1\over2},0,0,0,-{1\over2}}&(3c)\cr}$$

There are several sectors that can, a priori, produce electroweak
Higgs doublets and color triplets. The first is the Neveu--Schwarz sector.
The Neveu--Schwarz sector correspond to the untwisted sector in the
orbifold formulation. At the level of the NAHE set the Neveu--Schwarz
produces three vector $10$ representation of $SO(10)$, which decompose
as $5+\bar 5$ under $SU(5)$. These are obtained by acting on the vacuum with
${\chi_{1\over2}^{j}}{{\bar\psi}_{1\over2}^{{1\cdots5}}}
{\bar\eta}^j_{1\over2}$, $(j=1,2,3)$,
where $\chi^j_{1\over2}$, ($j=1,2,3$), denotes one of the pairs
$\chi^{12}_{1\over2},\chi^{34}_{1\over2},\chi^{56}_{1\over2}$
respectively.

The second sector that may produce color triplet supermultiplets is the sector
$\zeta=b_1+b_2+\alpha+\beta$. The unique property of this vector
is that it does not have periodic world--sheet fermions under
$SO(10)\times E_8$ and it has
$\zeta_R\cdot{\zeta_R}=\zeta_L\cdot{\zeta_L}=4$.
The massless states are obtained by acting on the vacuum
with one right--moving fermionic oscillator. The states in this sector
transform only under the observable gauge group. The vector
combination that produces these additional color triplets is not generic
to all the free fermionic models. For example, in the model of Ref. [\FNY]
such a vector combination does not exist in the additive group.
However, requiring realistic
low energy mass spectrum necessitates the presence of such a vector
combination in the additive group [\SLM,\NRT].

The Neveu--Schwarz sector and the sector
$\zeta=b_1+b_2+\alpha+\beta$ may produce
the most dangerous Higgs color triplet representations. The reason is that
the states from these two sectors transform solely under the observable
gauge group. In addition to these two sectors, two additional
type of sectors may produce additional color triplets. These
sectors are obtained from combination of the vectors
$\{b_1,b_2,b_3,\alpha,\beta\}\pm\gamma$ or $+2\gamma$. The two type
of sectors are distinguished by the product $\alpha_R\cdot\alpha_R=6,8$.
For the first type, $\alpha_R\cdot\alpha_R=6$ and the color triplets
are obtained by acting on the vacuum with a right--moving
fermionic oscillator, while in the second type, the vacuum is composed
entirely of Ramond vacua. Table 1 summarizes the additional
color triplets in the model of Ref. [\GCU].

\bigskip
\centerline{\bf 3.  The superstring doublet--triplet splitting mechanism}

The additional vectors $\{\alpha,\beta,\gamma\}$ break the gauge
symmetry and reduce the number of generations to three. The final
observable and hidden gauge groups depend on the assignment of
boundary conditions, in these vectors, to the sets $\{{\bar\psi}^{1,\cdots,5}
,{\bar\eta}^{1,2,3}\}$ and $\{{\bar\phi}^{1,\cdots8}\}$, respectively.
The final number of generations depends on the assignment of boundary
conditions for the set of internal fermions
$\{y,\omega\vert{\bar y},{\bar\omega}\}^{1,\cdots,6}$. This set of internal
fermions corresponds to the six compactified dimension in a bosonic
formulation or to the left--right symmetric internal conformal field
theory. Whether or not the massless spectrum contains color Higgs triplets
depends as well on the assignment of boundary conditions
to the set of left--right symmetric internal fermions,
$\{y,\omega\vert{\bar y},{\bar\omega}\}^{1,\cdots,6}$.

First, I examine the $5+{\bar 5}$ representations from the
Neveu--Schwarz sector. In this case we observe a doublet--triplet
splitting mechanism that is correlated with the presence
of additional horizontal $U(1)$ symmetries that arise from the
set of left--right symmetric internal fermions. The additional
$U(1)$ symmetries arise from pairing real right--moving fermions
to form a complex fermion. Alternatively a right--moving real
fermion can be paired with a left--moving real fermion to form an
Ising model operator [\KLN].
The assignment of boundary conditions in all vectors is identical
for these pairs, and there exists at least one vector combination
that separates them from all the other internal fermions.
For every right--moving pair that produces a gauged $U(1)$ current
there is a symmetric left--moving
pair that produces a global $U(1)$ symmetry.
These pairs of right and left--moving internal fermions
guarantee that the color triplets, $D_j$ and ${\bar D}_j$
from the Neveu--Schwarz sector are
projected out and that the
Higgs doublets, $h_j$ and $\bar h_j$,
remain in the massless spectrum.
A selection rule is observed in the application
of the GSO projection, $\alpha$, which breaks the  $SO(10)$ symmetry to
$SO(6)\times SO(4)$.
I denote by $\alpha_L(b_j)$, and $\alpha_R(b_j)$, the intersection of
the periodic boundary conditions , in the vectors
$\alpha$ and $b_j$, of the sets of internal fermions
$\{y,\omega\}_L$, and $\{{\bar y},{\bar\omega}\}_R$, respectively.
The superstring doublet--triplet selection rule
then says:

If $\vert\alpha_L(b_j)-\alpha_R(b_j)\vert=0$ then the electroweak
doublets, $h_j$, are projected out and the color triplets, $D_j$,
remain in the physical spectrum. if $\vert\alpha_L(b_j)-\alpha_R(b_j)\vert=1$
then the color triplets are projected out and the electroweak doublets
remain in the physical spectrum.

I now prove this doublet--triplet selection rule. The only states that
contribute to the physical spectrum are those that satisfy the
generalized GSO projections,
$$\left\{e^{i\pi({b_i}F_\alpha)}-
{\delta_\alpha}c^*\left(\matrix{\alpha\cr
                 b_i\cr}\right)\right\}\vert{s}\rangle=0\eqno(4a)$$
with $$(b_i{F_\alpha})\equiv\{\sum_{real+complex\atop{left}}-
\sum_{real+complex\atop{right}}\}(b_i(f)F_\alpha(f)),\eqno(4b)$$
where $F_\alpha(f)$ is a fermion number operator counting each mode of
$f$ once (and if $f$ is complex, $f^*$ minus once). For periodic
fermions the vacuum is a spinor in order to represent
the Clifford algebra of the corresponding zero modes.
For each periodic complex fermion $f$,
there are two degenerate vacua $\vert{+}\rangle$, $\vert{-}\rangle$,
annihilated by the zero modes $f_0$ and $f^*_0$ and with fermion
number $F(f)=0,-1$ respectively. In Eq. (7a),
$\delta_\alpha=-1$ if $\psi^\mu$ is periodic in the sector $\alpha$,
and $\delta_\alpha=+1$ if $\psi^\mu$ is antiperiodic in the sector $\alpha$.

The $5+{\bar 5}$ representations from the Neveu--Schwarz
sector are of the form,
$$h_j\equiv{\chi^j_{1\over2}{\bar\psi}^{{1,\cdots,5}^*}_{1\over2}
{\bar\eta}^j_{1\over2}},\eqno(5)$$
The GSO projection coefficients for the Neveu--Schwarz sector
are $$c^*\left(\matrix{NS\cr
                 \alpha\cr}\right)=\delta_{\alpha}=
exp(i\pi\alpha(\psi^\mu))\eqno(6)$$ and $\delta_{NS}=+1$.
By operating
 the GSO projection, Eq. (4), of the vector $\alpha$ on the
 representations, Eq. (5), the following equation is obtained,
$$\alpha(\chi_j)+\alpha({\bar\psi}^{1,\cdots,3})+\alpha({\bar\psi}^{4,5})+
\alpha({\bar\eta}_j)=\alpha(\psi^\mu)~mod~2.\eqno(7)$$
The modular invariance constraint on the product of two basis vectors
gives
$$\alpha\cdot{b_j}=\alpha(\psi^\mu)+\alpha(\chi^j)+
(\alpha_L(b_j)-\alpha_R(b_j))
-\sum_{i=1}^{3}\alpha({\bar\psi}^{1,\cdots,3})-
\alpha({\bar\eta}_j)=0~mod~2 ,\eqno(8)$$ where
$(\alpha_L(b_j)-\alpha_R(b_j))$
is the difference of the left--right symmetric part of
$\alpha\cdot b_j$.
In the vector $\alpha$, $\alpha({\bar\psi}^{1,\cdots,3})=1$
and $\alpha({\bar\psi}^{4,5})=0$.
There are four possible boundary conditions for the pair $(\psi^\mu,\chi^j)$
in the vector $\alpha$, $\alpha(\psi^\mu,\chi^j)=\{(1,1);(1,0);(0,1);(0,0)\}$.
I take, for example, the first case $\alpha(\psi^\mu,\chi^j)=(1,1)$.
In this case, for $\vert(\alpha_L(b_j)-\alpha_R(b_j))\vert=0$,
Eqs. (7,8) reduce to
$$\alpha({\bar\psi}^{1,\cdots,3})+\alpha({\bar\psi}^{4,5})+
\alpha({\bar\eta}_j)=0~mod~2.\eqno(9)$$
and
$$-\sum_{i=1}^{3}\alpha({\bar\psi}^{1,\cdots,3})-
\alpha({\bar\eta}_j)=0~mod~2.\eqno(10)$$
Because ${\sum_{i=1}^{3}}\alpha({\bar\psi}^{1,\cdots,3})=3$,
from Eq. (10) follows $\alpha({\bar\eta}_j)=1$. Then Eq. (9) can be satisfied
for the triplets but cannot be satisfied for the doublets.
Therefore the electroweak doublets are projected out while the color
triplets remain in the massless spectrum.

On the other hand, for $\vert(\alpha_L(b_j)-\alpha_R(b_j))\vert=1$,
Eq. (8) reduces to
$$-\sum_{i=1}^{3}\alpha({\bar\psi}^{1,\cdots,3})-
\alpha({\bar\eta}_j)=1~mod~2 ,\eqno(11)$$
and with $\alpha(\psi^\mu,\chi^j)=(1,1)$
Eq. (7) remains as in Eq. (9). Therefore in this case, from Eq. (11) follows
$\alpha({\bar\eta}_j)=0$. In this case
the triplets cannot satisfy Eq. (9) and are therefore projected out, while
the doublets remain in the massless spectrum.
In a similar way this selection rule can be shown to hold
for the other choices of boundary conditions, in the vector $\alpha$,
for the pair $(\psi^\mu,\chi^j)$.
To summarize, the constraint
$$\vert\alpha_L(b_j)-\alpha_R(b_j)\vert=1~~,~~(j=1,2,3), \eqno(12)$$
guarantees that the Neveu--Schwarz color triplets, $D_j$,
are projected out and that
the electroweak doublets remain in the massless spectrum.

To illustrate this dependence I consider the models in tables  2, 3, 4
and 5.
In the models of tables 2 and 3, the three horizontal ($U(1)_\ell;U(1)_r$)
symmetries, which correspond  to the
world-sheet currents $(y^3y^6;{\bar y}^3{\bar y}^6)$,
$(y^1\omega^5;{\bar y}^1{\bar\omega}^5)$
and $(\omega^2\omega^4;{\bar\omega}^2{\bar\omega}^4)$,
guarantee that the Higgs doublets $h_1$, ${\bar h}_1$, $h_2$, ${\bar h}_2$
and $h_3$, ${\bar h}_3$ remain in the massless spectrum, and that the
color triplets are projected out.
In the model of table 4 all the real fermions are paired to form
Ising model operators and there are no additional $U(1)$ symmetries
beyond $U(1)_{r_j}$ $(j=1,2,3)$. All the Higgs doublets from the
Neveu--Schwarz sector are projected out. In this case the Higgs
triplets $D_1$, $\bar D_1$, $D_2$, $\bar D_2$ and $D_3$, $\bar D_3$
remain in the massless spectrum.
In model 5 we have only one additional horizontal
($U(1)_\ell;U(1)_r$) symmetry which corresponds to the world--sheet currents
$(\omega^2\omega^3;{\bar\omega^2}{\bar\omega^3})$.
Therefore in this model
only one pair of Higgs doublets from the Neveu--Schwarz sector,
$h_3$, ${\bar h}_3$, remains in
the massless spectrum after the GSO projections. In this case we obtain
the color triplets $D_1$, $\bar D_1$ and $D_2$, $\bar D_2$.

Next I turn to the sector $S+b_1+b_2+\alpha+\beta$. This sector
produces additional states that transform solely under the observable
sector. In particular it can give rise to additional electroweak
doublets and color triplets. To obtain realistic low energy
phenomenology, it is essential that such a vector
combination exists in the additive group.
The color triplets from this sector may cause problems with proton
lifetime constraints.
However, a similar
doublet--triplet splitting mechanism works for this sector as well.
There exist choices of boundary conditions for the set of
left--right symmetric internal fermions,
$\{y,\omega\vert{\bar y},{\bar\omega}\}^{1,\cdots,6}$, for which
the triplets are projected out and the doublets remain in the massless
spectrum. For example, in the model of Ref. [\EU] (table 2)
this sector produces
one pair of electroweak doublets and one pair of color triplets.
$$\eqalignno{h_{45}&\equiv{[(1,0);(2,-1)]}_
{-{1\over2},-{1\over2},0,0,0,0} {\hskip .3cm}
D_{45}\equiv{[(3,-1);(1,0)]}_
{-{1\over2},-{1\over2},0,0,0,0}&(13a,b)\cr}$$
while in the model of Ref. [\TOP,\GCU] (table 3) this sector produces two
pairs of electroweak doublets,
$$\eqalignno{h_{45}&\equiv{[(1,0);(2,-1)]}_
{{1\over2},{1\over2},0,0,0,0} {\hskip .4cm}
{h}_{45}^\prime\equiv{[(1,0);(2,-1)]}_
{-{1\over2},-{1\over2},0,0,0,0}&(14a,b)\cr}$$
and all the color triplets, from the Neveu--Schwarz sector and the sector
$b_1+b_2+\alpha+\gamma$, are projected
from the physical spectrum by the GSO projections. As is
evident from tables 2 and 3, the two models differ only by the assignment
of boundary conditions to the set of internal fermions,
$\{y,\omega\vert{\bar y},{\bar\omega}\}^{1,\cdots,6}$.
The simplicity and elegance of the superstring doublet--triplet
splitting mechanism is striking. There is no need for exotic
representations of high dimensionality as in minimal $SU(5)$ extension
of the Standard Model [\G]. Moreover, the superstring
doublet--triplet splitting mechanism does not depend on additional
assumptions on Yukawa couplings as is required in all GUT
doublet--triplet splitting mechanism. In the superstring
doublet--triplet splitting mechanism the dangerous color triplets
simply do not exist in the massless spectrum.
In the next section I investigate the implications on proton decay.

\bigskip
\centerline{\bf 4. Proton decay}

In the most general supersymmetric standard model
the dimension four operators,
${\eta_1}{u_{L}^C}{d_{L}^C}{d_{L}^C}+
{\eta_2}{d_{L}^C}QL$, mediate instantaneous proton decay if $\eta_1$ and
$\eta_2$ are both large.
Traditionally in supersymmetric models, one imposes R symmetries on the
spectrum to avoid this problem. In the context of superstring theories
these discrete symmetries are usually not found [\FIQ].
These dimension
four operators are forbidden if the gauge symmetry of the
Standard Model is extended by a
$U(1)$ symmetry, which
is a combination of, $B-L$, baryon
minus lepton number,  and $T_{3_R}$, and is exactly the
additional, generation independent, $U(1)$ symmetry
that is derived in the superstring standard--like models.
However, the dimension four operators may still
appear from nonrenormalizable terms.
In terms of $SO(10)$ representations the effective dimension
four operators are obtained from a
$16^4$ operator, if one of the $16$ obtains a VEV.
In terms of standard model multiplets the dangerous operators are
$${\eta_1}({u_{L}^C}{d_{L}^C}{d_{L}^C}N_L^c)\Phi+
{\eta_2}({d_{L}^C}QLN_L^c)\Phi,\eqno(15)$$
where $N_L^c$ is the Standard Model singlet in the $16$ of $SO(10)$.
$\Phi$ is a combination of fields which
fixes the string selection rules and gets a VEV of $O(M/10)$, where
$M=M_{Pl}/2\sqrt{8\pi}$. From Eq. (15) it is seen that the ratio
${\langle{N_L^c}\rangle}/{M}$ controls the rate
of proton decay. While in the standard--like models we may impose
$\l N_L^c\r\equiv0$, or $\l N_L^c\r$ smaller than some value,
in superstring models that are based on an intermediate GUT symmetry
the problem is more acute as $\l N_L^c\r$ is necessarily used to
break the GUT symmetry. A search through nonrenormalizable
terms shows that terms of the form of Eq. (15) are in general
generated in string models [\FBV,\NRT].

To guarantee that the dimension four operators are sufficiently
suppressed we must ascertain that other $U(1)_{Z^\prime}$
breaking VEVs cannot generate them. In addition to $N_L^c$ the massless
spectrum of the superstring standard--like models contain neutral states
with fractional $U(1)_{Z^\prime}$. To produce the $U(1)_{Z^\prime}$ charge
of $N_L^c$ two such states have to be combined.
The possible terms must have the form
$$u_L^cd_L^cd_L^cHH\phi^n~~\hbox{and}~~QLd_L^cHH\phi^n\eqno(16)$$
where $\l H\r$ breaks  $U(1)_{Z^\prime}$ and
$\phi^n$ is a string of Standard Model singlets that fixes the
remaining string selection rules. For example, in the model of Ref. [\GCU]
all the states with fractional $U(1)_{Z^\prime}$ charge transform as
$3$ and ${\bar 3}$ of the hidden $SU(3)$ group  (see table 1).
Therefore, in this model
invariance under the hidden Abelian and non--Abelian gauge groups,
and under $U(1)_{Z^\prime}$,
forbid the formation
of terms of the form of Eq. (16) to all orders of nonrenormalizable terms.
However, in general, such terms can be formed. For example in the model
of Ref. [\EU] such terms appear at order $N=8$ and will
not be enumerated here. They contain a suppression factor of
$(\Lambda_{Z^\prime}/M)^2$ and are therefore sufficiently suppressed if,
for example, $\Lambda_{Z^\prime}\le10^{12}~GeV$.

Next I turn to proton decay from dimension five operators.
The dangerous dimension five operators are:
$$QQQh,~~~~~QQQL,~~~~~d_L^cu_L^cu_L^ce_L^c\eqno(17)$$
The first operator is forbidden by $U(1)_{Z^\prime}$ charge conservation.
Tagging to it ${\bar N}_L^c$, from the ${\bar{16}}$ of $SO(10)$ renders
it invariant under $U(1)_{Z^\prime}$. In the superstring standard--like models
the ${\bar{16}}$ of $SO(10)$ is not present in the massless spectrum.
One may still form combination states of states with fractional
$U(1)_{Z^\prime}$ charge that effectively produce the same charge as
${\bar N}_L^c$. However, this introduces an additional
suppression by $\Lambda_{Z^\prime}/M$. Therefore, this operators is
suppressed by at least $\Lambda_{Z^\prime}^2/M^3$. Thus, even for
rather large $\Lambda_{Z^\prime}\sim10^{15}~GeV$, this operator is
harmless.
Therefore, in the superstring standard--like models
there are two, potentially dangerous,
Baryon violating dimension five operators, $QQQL$ and $d_L^cu_L^cu_L^ce_L^c$.
The second operator does not contribute to proton decay [\SO]. Therefore,
we are left with a single operator, $QQQL$.

There are two possible ways in which these operators could be
generated. One is through triplet Higgsino exchange.
In this case the dimension five operators are obtained from the tree
level operators
$$LQ{\bar D},~u_L^ce_L^cD,~QQD,~u_L^cd_L^c{\bar D},~d_L^cN_L^cD,
{}~D{\bar D}\phi\eqno(18)$$
where $D$ are the color triplets in the $5$ of $SU(5)$.
Alternatively the dimension five operators may be induced by
exchange of heavy string modes. To insure that dangerous dimension five
operators are not induced we must check that both the terms in Eq. (18)
are suppressed and that the dimension five operators are not
induced from nonrenormalizable terms at the string level.
In general, if massless triplets from the Neveu--Schwarz sector
or the sector $b_1+b_2+\alpha+\beta$  exist in the massless spectrum then
then the terms in Eq. (18) are obtained either at the cubic
level of the superpotential or from higher order nonrenormalizable
terms. For example in the model of table 4
(the massless
spectrum and quantum numbers are given in Ref. [\SLM]),
we obtain at the cubic level,
$$\eqalignno{&
{u_{L_1}^c}{e_{L_1}^c}{D}_1,~{d_{L_1}^c}{N_{L_1}^c}{D}_1,~
{u_{L_2}^c}{e_{L_2}^c}{D}_2,~{d_{L_1}^c}{N_{L_2}^c}{D}_1,\cr
&\qquad
{{D_1}{\bar D}_2{\bar\Phi}_{12}},~
{\bar D}_1{D}_2{\Phi}_{12}\cr}$$
while in the model of Ref. [\EU] we obtain at the quartic order,
$$\eqalignno{&Q_1Q_1D_{45}\Phi_1^+,~
              Q_2Q_2D_{45}{\bar\Phi}_2^-,~
              u_1^ce_1^c{\bar D}_{45}{\bar\Phi}^-_1,~
              u_2^ce_2^cD_{45}{\bar\Phi}^+_2,~
              d_1^cN_1^cD_{45}\Phi_1^+,~
              d_2^cN_2^cD_{45}{\bar\Phi}_2^-,~&\cr
             &u_2^ce_2^cH_{21}H_{26},~
              Q_2L_2H_{21}H_{26},\cr}$$
In general, it is expected that any term that is not forbidden by some
symmetry will be generated at some order and therefore may cause
problems with proton decay. The validity of the models that
contain the dangerous Higgs triplets then rests upon the ability to find
flat directions that suppress the dangerous operators.
This approach is not very appealing as the choice of flat directions
depends also on other phenomenological requirements. Furthermore,
if the color triplets are not sufficiently heavy the the mixing between
the two Higgs triplets must be suppressed. Otherwise, effective
dimension five operators may result from $D$ and ${\bar D}$ exchange.

In the superstring standard--like models
there exist a more elegant solution to the problem with
dimension five operators from Higgsino exchange. The color triplets
that may couple to the Standard Model multiplets are projected out from the
massless spectrum by means of the GSO projections. There are no dimension
five operators from Higgsino exchange, simply because there are no
color Higgs triplets that can produce them. The decisiveness and
elegance
of the superstring doublet--triplet splitting mechanism
can only be highlighted when compared to the proposed
solutions in GUT models. In the minimal $SU(5)$ the doublet--triplet
splitting mechanism rests upon the existence of very large representations,
and then assumes a potential that insures that all the extra matter
becomes supermassive. In the flipped $SU(5)$ model
there is an elegant doublet--triplet splitting in which
the Higgs triplets receive
GUT scale mass by coupling them to the triplets in the representations
that are used to break the GUT symmetry. These representations are
$10$ and $\bar{10}$ of $SU(5)$ that are embedded in the $16$ and $\bar{16}$
of $SO(10)$. The couplings
$\lambda_4HHh+\lambda_5{\bar H}{\bar H}{\bar h}$ then give large Dirac masses
to $D$ and ${\bar D}$.
However, also there the mechanism depends on additional assumptions
on the coupling $\lambda_4$ and $\lambda_5$. Therefore,
more assumptions are made. In contrast, in the superstring standard--like
models no such assumptions are needed. The dangerous Higgs color
triplets are simply not there.

I now examine the dimension five operators in the models of
Ref. [\TOP,\GCU]. In these models the triplets from the
Neveu--Schwarz as well as the triplets from the sector
$b_1+b_2+\alpha+\beta$ are projected out from the
physical spectrum due to the assignment
of boundary conditions in the vectors $\alpha$, $\beta$
and $\gamma$. The sector $b_1+b_2+\alpha+\beta$
produces two pairs of electroweak doublets rather than a pair of
triplets and a pair of doublets. Thus, in this subclass of
superstring standard--like models, there are no color triplets
from the Neveu--Schwarz sector, nor from the sector
$b_1+b_2+\alpha+\beta$. We have also to examine the interactions of
the Standard Model states with additional triplets in the massless spectrum.
The models of Ref. [\TOP,\GCU] contain additional
triplets from sectors that arise due to Wilson
line breaking. The number of such triplets is maximized in the model
of Ref. [\GCU]. Therefore, in what follows, I focus on this model.

The additional triplets and their quantum numbers under the right--moving
$U(1)$ symmetries are given in table 1. The type of
correlators that have to be checked are of the form
$b_ib_j{D}\phi^n$, where $b_i$ and $b_j$ represent states from
the sectors $b_i$ and $b_j$, $D$ are the additional color triplets,
and $\phi^n$ is a string of Standard Model singlets.
For the first two pairs of color triplets from the sectors
$b_{1,2}+b_3+\alpha+\beta$,
the operators $b_ib_j{D}$ are invariant under the weak hypercharge.
However, they break $U(1)_{Z^\prime}$ because
$Q_{Z^\prime}({D})={(1/2)}Q_{Z^\prime}({D_{45}})$.
Thus, $D$ has one half the $U(1)_{Z^\prime}$ charge of the triplets
from the Neveu--Schwarz and $b_1+b_2+\alpha+\beta$ sectors.
Therefore, all the operators in Eq. (9), with $D$ being a triplet from one of
the sectors $b_{1,2}+b_3+\alpha+\beta$, break $U(1)_{Z^\prime}$.
Thus, the string $\l\phi\r^n$
contains a $U(1)_{Z^\prime}$ breaking VEV. However, in these model
all the available
Standard Model singlets with nontrivial $U(1)_{Z^\prime}$ charge
transform as $3$ and ${\bar 3}$ of the Hidden $SU(3)$ gauge group
(see table 1).
The $U(1)_{Z^\prime}$ charges of the hidden $SU(3)$ triplets are
$\pm5/4$. The $U(1)_{Z^\prime}$ charges of the color triplets from the
exotic ``Wilson line'' sectors are $\pm1/4$ (see table 1).
The last pair of color triplets has ``fractional'' weak hypercharge
$Q_Y=\pm1/12$.
Therefore, terms of the form of Eq. (18),
with $D$ being a triplet from one of
the exotic ``Wilson line'' sectors, cannot be formed in this model.
Therefore, in this model proton decay from dimension five or six operators
due to Higgsino or Higgs exchange is suppressed.

Next I show  that the dimension five operators cannot be
generated by exchange of heavy string modes. I show that the
dangerous dimension five operators are forbidden to all orders
of nonrenormalizable terms. This follows from the charges of the states
from the sectors $b_j$ under the horizontal $U(1)_{r_{1,\cdots,6}}$
(see Eq. (1)). All the states from the sectors $b_j$ have charge
$1/2$ under $U(1)_{r_j}$ while under $U(1)_{r_{j+3}}$
we have
$$U(1)_{r_{j+3}}(Q,L)=-{1\over2}~~\hbox{and}~~
U(1)_{r_{j+3}}(d_L^c,u_L^c,e_L^c)={1\over2}$$
Consider the symmetry $U(1)^\prime=U(1)_4+U(1)_5+U(1)_6$.
Under this symmetry, $U(1)^\prime(Q_j)=-1/2$ and $U(1)^\prime(L_j)=-1/2$.
The charge of the correlator $QQQL$ under $U(1)^\prime$ is $-2$. The
only additional states in the massless spectrum with, $U(1)^\prime\ne0$ are
the states from the sectors $b_j+2\gamma$.
However, $$U(1)^\prime(3_j)=-U(1)^\prime({\bar3}_j)~~\hbox{and}~~
U(1)^\prime(5_j)=-U(1)^\prime({\bar5})_j~.$$
In any correlator we can only have
$\l3{\bar 3}\r$ or $\l5{\bar5}\r$, to respect invariance under the Abelian and
non--Abelian hidden gauge groups. Therefore the correlator $QQQL$
cannot be invariant under $U(1)^\prime$. Similarly, the total charge
of $d_L^cu_L^cu_L^ce_L^c$
under $U(1)^\prime$ is $+2$. Therefore the correlator
$d_L^cu_L^cu_L^ce_L^c$ also cannot be invariant under $U(1)^\prime$.
This completes the proof that there no dimension five operators in this
model.
{}From similar considerations it is also seen that, while the
operator $QLd_L^cN_L^c$ is allowed, the operator $u_L^cd_L^cd_L^cN_L^c$
is forbidden by the $U(1)^\prime$ symmetry. Therefore, in this model
there is no proton decay from dimension four, five and six operators.

To summarize, in the case of the standard--like models that contain color
triplets from the Neveu--Schwarz or $b_1+b_2+\alpha+\beta$ sectors,
the problem with proton decay is similar to the problem encountered
in traditional SUSY GUT models. One must find specific
choices of flat directions for which the dangerous operators are
suppressed. However, there is a class of superstring
standard--like models in which the color triplets from the
Neveu--Schwarz sector and from the sector $b_1+b_2+\alpha+\beta$
are projected out by the GSO projections. In this class of models,
proton decay from triplet Higgsino or triplet Higgs exchange is
forbidden. The right--moving gauged $U(1)$ symmetries forbid the
formation of dimension five operators from nonrenormalizable terms,
to all orders. Moreover, the same symmetries also forbid proton decay
from effective dimension four operators to all orders of nonrenormalizable
terms. Thus, in this class of models there cannot be proton decay
from dimension four, five and six operators, to all
orders of nonrenormalizable terms.

\refout
\vfill\eject

\input tables.tex
\nopagenumbers
\magnification=1000
\baselineskip=18pt
\hbox
{\hfill
{\begintable
\ F \ \|\ SEC \ \|\ $SU(3)_C$ $\times$ $SU(2)_L$ \ \|\ $Q_C$ & $Q_L$ & $Q_1$ &
$Q_2$
 & $Q_3$ & $Q_4$ & $Q_5$ & $Q_6$ \ \|\ $SU(5)$ $\times$ $SU(3)$ \ \|\ $Q_7$ &
$Q_8$  \crthick
$D_1$ \|\ ${b_2+b_3+\beta}$ \|(3,1)\|~~$1\over4$ & ~~$1\over2$
 & ~~${1\over4}$ & $-{1\over 4}$ &
 ~~${1\over 4}$ & ~~0 & ~~0 & ~~0 \|(1,1)\| $-{1\over4}$ &
$-{{15}\over 4}$   \nr
${\bar D}_1$ \|\ ${\pm\gamma}+(I)$
\|(${\bar 3}$,1)\| $-{1\over4}$ &
$-{1\over2}$ & $-{1\over4}$ & ~~${1\over 4}$ &
 ~~${1\over 4}$ & ~~0 & ~~0 & ~~0
\|(1,1)\| ~~${1\over 4}$ & ~~${{15}\over 4}$  \cr
$D_2$ \|\ ${b_1+b_3+\alpha}$ \|(3,1)\|~~$1\over4$ & ~~$1\over2$
 & $-{1\over4}$ & ~~${1\over 4}$ &
 $-{1\over 4}$ & ~~0 & ~~0 & ~~0 \|(1,1)\| $-{1\over 4}$ &
$-{{15}\over 4}$   \nr
${\bar D}_2$ \|\ ${\pm\gamma}+(I)$
\|(${\bar 3}$,1)\| $-{1\over4}$ & $-{1\over2}$
& ~~${1\over4}$ & $-{1\over 4}$ &
 ~~${1\over 4}$ & ~~0 & ~~0 & ~~0
\|(1,1)\| ~~${1\over 4}$ & ~~${{15}\over 4}$  \cr
$D_3$ \|\ ${1+\alpha}$ \|(3,1)\|~~$1\over2$ & ~~0
 & ~~0 & ~~0 &
 ~~0 & ~~${1\over 2}$ & ~~${1\over2}$ &
{}~~${1\over2}$ \|(1,1)\| $-1$ &
{}~~0   \nr
${\bar D}_3$ \|\ $+2\gamma$
\|(${\bar 3}$,1)\| $-{1\over2}$ & ~~0 & ~~0 & ~~0 &
 ~~0 & $-{1\over 2}$ & $-{1\over 2}$ & ~~${1\over2}$
\|(1,1)\| ~~1 & ~~0  \crthick
$H_1$ \|\ ${b_2+b_3+\beta}$ \|(1,1)\| $-{3\over4}$ & ~~$1\over2$
 & $-{1\over4}$ & ~~${1\over 4}$ &
 $-{1\over 4}$ & ~~0 & ~~0 & ~~0 \|(1,3)\| ~~${3\over4}$ &
{}~~${{5}\over 4}$   \nr
${\bar H}_1$ \|\ $\pm\gamma+(I)$
\|(1,1)\| ~~${3\over4}$ &
$-{1\over2}$ & ~~${1\over4}$ & $-{1\over 4}$ &
 ~~${1\over 4}$ & ~~0 & ~~0 & ~~0
\|(1,${\bar3}$)\| $-{3\over 4}$ & $-{{5}\over 4}$  \cr
$H_2$ \|\ ${b_1+b_3+\alpha}$ \|(1,1)\| $-{3\over4}$ & ~~${1\over2}$
 & ~~${1\over4}$ & $-{1\over 4}$ &
 $-{1\over 4}$ & ~~0 & ~~0 & ~~0 \|(1,3)\| ~~${3\over 4}$ &
{}~~${{5}\over 4}$   \nr
${\bar H}_2$ \|\ $\pm\gamma+(I)$
\|(1,1)\| ~~${3\over4}$ & $-{1\over2}$
& $-{1\over4}$ & ~~${1\over 4}$ &
 ~~${1\over 4}$ & ~~0 & ~~0 & ~~0
\|(1,${\bar 3}$)\| $-{3\over 4}$ & $-{{5}\over 4}$
\endtable}
\hfill}
\bigskip
\parindent=0pt
\hangindent=39pt\hangafter=1
{\it Table 1.}
Massless states and their quantum numbers in the model of table 3.
The first three pairs are additional color triplets. The last
two pairs are the states with vanishing weak hypercharge and fractional
$U(1)_{Z^\prime}$ charge.

\vfill
\eject
\magnification=1000
\tolerance=1200



{\hfill
{\begintable
\  \ \|\ ${\psi^\mu}$ \ \|\ $\{{\chi^{12};\chi^{34};\chi^{56}}\}$\ \|\
{${\bar\psi}^1$, ${\bar\psi}^2$, ${\bar\psi}^3$,
${\bar\psi}^4$, ${\bar\psi}^5$, ${\bar\eta}^1$,
${\bar\eta}^2$, ${\bar\eta}^3$} \ \|\
{${\bar\phi}^1$, ${\bar\phi}^2$, ${\bar\phi}^3$, ${\bar\phi}^4$,
${\bar\phi}^5$, ${\bar\phi}^6$, ${\bar\phi}^7$, ${\bar\phi}^8$} \crthick
$\alpha$
\|\ 0 \|
$\{0,~0,~0\}$ \|
1, ~~1, ~~1, ~~0, ~~0, ~~0 ,~~0, ~~0 \|
1, ~~1, ~~1, ~~1, ~~0, ~~0, ~~0, ~~0 \nr
$\beta$
\|\ 0 \| $\{0,~0,~0\}$ \|
1, ~~1, ~~1, ~~0, ~~0, ~~0, ~~0, ~~0 \|
1, ~~1, ~~1, ~~1, ~~0, ~~0, ~~0, ~~0 \nr
$\gamma$
\|\ 0 \|
$\{0,~0,~0\}$ \|
{}~~$1\over2$, ~~$1\over2$, ~~$1\over2$, ~~$1\over2$,
{}~~$1\over2$, ~~$1\over2$, ~~$1\over2$, ~~$1\over2$ \| $1\over2$, ~~0, ~~1,
{}~~1,
{}~~$1\over2$,
{}~~$1\over2$, ~~$1\over2$, ~~0 \endtable}
\hfill}
\smallskip
{\hfill
{\begintable
\  \ \|\
${y^3y^6}$,  ${y^4{\bar y}^4}$, ${y^5{\bar y}^5}$,
${{\bar y}^3{\bar y}^6}$
\ \|\ ${y^1\omega^6}$,  ${y^2{\bar y}^2}$,
${\omega^5{\bar\omega}^5}$,
${{\bar y}^1{\bar\omega}^6}$
\ \|\ ${\omega^1{\omega}^3}$,  ${\omega^2{\bar\omega}^2}$,
${\omega^4{\bar\omega}^4}$,  ${{\bar\omega}^1{\bar\omega}^3}$  \crthick
$\alpha$ \|
1, ~~~0, ~~~~0, ~~~~0 \|
0, ~~~0, ~~~~1, ~~~~1 \|
0, ~~~0, ~~~~1, ~~~~1 \nr
$\beta$ \|
0, ~~~0, ~~~~1, ~~~~1 \|
1, ~~~0, ~~~~0, ~~~~0 \|
0, ~~~1, ~~~~0, ~~~~1 \nr
$\gamma$ \|
0, ~~~1, ~~~~0, ~~~~1 \|\
0, ~~~1, ~~~~0, ~~~~1 \|
1, ~~~0, ~~~~0, ~~~~0  \endtable}
\hfill}
\smallskip
\parindent=0pt
\hangindent=39pt\hangafter=1
\baselineskip=18pt

{{\it Table 2.} A three generations
${SU(3)\times SU(2)\times U(1)^2}$ model,
with $\vert\alpha_L(b_j)-\alpha_R(b_j)\vert=1~~,~~(j=1,2,3)$.
The color triplets from the Neveu--Schwarz sector
are projected out and the electroweak triplets remain in
the physical spectrum. The sector $b_1+b_2+\alpha+\beta$ produces
one pair of electroweak doublets and one pair of color triplets.
The $16$ right--moving
internal fermionic states
$\{{\bar\psi}^{1,\cdots,5},{\bar\eta}^1,
{\bar\eta}^2,{\bar\eta}^3,{\bar\phi}^{1,\cdots,8}\}$,
correspond to the $16$ dimensional compactified  torus of the ten dimensional
heterotic string. The 12 left--moving and 12 right--moving
real internal fermionic states
correspond to the six left and six right compactified
dimensions in the bosonic language.
$\psi^\mu$ are the two space--time external fermions
in the light--cone gauge and
$\chi^{12}$, $\chi^{34}$, $\chi^{56}$ correspond to the
spin connection in the bosonic
constructions.}

\vskip 2.5cm

\vfill
\eject

\magnification=1000
\tolerance=1200


{\hfill
{\begintable
\  \ \|\ ${\psi^\mu}$ \ \|\ $\{{\chi^{12};\chi^{34};\chi^{56}}\}$  \ \|\
{${\bar\psi}^1$, ${\bar\psi}^2$, ${\bar\psi}^3$,
${\bar\psi}^4$, ${\bar\psi}^5$, ${\bar\eta}^1$,
${\bar\eta}^2$, ${\bar\eta}^3$} \ \|\
{${\bar\phi}^1$, ${\bar\phi}^2$, ${\bar\phi}^3$, ${\bar\phi}^4$,
${\bar\phi}^5$, ${\bar\phi}^6$, ${\bar\phi}^7$, ${\bar\phi}^8$} \crthick
${\alpha}$
\|\ 0 \|
$\{0,~0,~0\}$  \|
1, ~~1, ~~1, ~~0, ~~0, ~~0, ~~0, ~~0 \|
1, ~~1, ~~1, ~~1, ~~0, ~~0, ~~0, ~~0 \nr
${\beta}$
\|\ 0 \| $\{0,~0,~0\}$  \|
1, ~~1, ~~1, ~~0, ~~0, ~~0, ~~0, ~~0 \|
1, ~~1, ~~1, ~~1, ~~0, ~~0, ~~0, ~~0 \nr
${\gamma}$
\|\ 0 \|
$\{0,~0,~0\}$  \|
 ~~$1\over2$, ~~$1\over2$, ~~$1\over2$, ~~$1\over2$,
{}~~$1\over2$, ~~$1\over2$, ~~$1\over2$, ~~$1\over2$ \|
$1\over2$, ~~0, ~~1, ~~1,
{}~~$1\over2$,
{}~~$1\over2$, ~~$1\over2$, ~~0 \endtable}
\hfill}
\smallskip
{\hfill
{\begintable
\  \ \|\
${y^3y^6}$,  ${y^4{\bar y}^4}$, ${y^5{\bar y}^5}$,
${{\bar y}^3{\bar y}^6}$
\ \|\ ${y^1\omega^6}$,  ${y^2{\bar y}^2}$,
${\omega^5{\bar\omega}^5}$,
${{\bar y}^1{\bar\omega}^6}$
\ \|\ ${\omega^1{\omega}^3}$,  ${\omega^2{\bar\omega}^2}$,
${\omega^4{\bar\omega}^4}$,  ${{\bar\omega}^1{\bar\omega}^3}$ \crthick
${\alpha}$ \|
1, ~~~1, ~~~~1, ~~~~0 \|
1, ~~~1, ~~~~1, ~~~~0 \|
1, ~~~1, ~~~~1, ~~~~0 \nr
${\beta}$ \|
0, ~~~1, ~~~~0, ~~~~1 \|
0, ~~~1, ~~~~0, ~~~~1 \|
1, ~~~0, ~~~~0, ~~~~0 \nr
${\gamma}$ \|
0, ~~~0, ~~~~1, ~~~~1 \|\
1, ~~~0, ~~~~0, ~~~~0 \|
0, ~~~1, ~~~~0, ~~~~1 \endtable}
\hfill}
\smallskip
\parindent=0pt
\hangindent=39pt\hangafter=1
\baselineskip=18pt
{{\it Table 3.} A three generations ${SU(3)\times SU(2)\times U(1)^2}$
model. The color triplets from the Neveu--Schwarz sector
are projected out and the electroweak triplets remain in
the physical spectrum. The color triplets from the
sector $b_1+b_2+\alpha+\beta$ are projected out as well.
The notation used is the notation of table 2. }

\vskip 2cm
\vfill
\eject

\magnification=1000
\tolerance=1200



{\hfill
{\begintable
\  \ \|\ ${\psi^\mu}$ \ \|\ $\{{\chi^{12};\chi^{34};\chi^{56}}\}$\ \|\
{${\bar\psi}^1$, ${\bar\psi}^2$, ${\bar\psi}^3$,
${\bar\psi}^4$, ${\bar\psi}^5$, ${\bar\eta}^1$,
${\bar\eta}^2$, ${\bar\eta}^3$} \ \|\
{${\bar\phi}^1$, ${\bar\phi}^2$, ${\bar\phi}^3$, ${\bar\phi}^4$,
${\bar\phi}^5$, ${\bar\phi}^6$, ${\bar\phi}^7$, ${\bar\phi}^8$} \crthick
$\alpha$
\|\ 1 \|
$\{1,~0,~0\}$ \|
1, ~~1, ~~1, ~~0, ~~0, ~~1 ,~~0, ~~0 \|
1, ~~1, ~~0, ~~0, ~~0, ~~0, ~~0, ~~0 \nr
$\beta$
\|\ 1 \|
$\{0,~1,~0\}$ \|
1, ~~1, ~~1, ~~0, ~~0, ~~0, ~~1, ~~0 \|
1, ~~1, ~~0, ~~0, ~~0, ~~0, ~~0, ~~0 \nr
$\gamma$
\|\ 1 \|
$\{0,~0,~1\}$ \|
{}~~$1\over2$, ~~$1\over2$, ~~$1\over2$, ~~$1\over2$,
{}~~$1\over2$, ~~$1\over2$, ~~$1\over2$, ~~$1\over2$ \|
$1\over2$, ~~$1\over2$, ~~$1\over2$, ~~$1\over2$, ~~1, ~~0, ~~0, ~~0 \endtable}
\hfill}
\smallskip
{\hfill
{\begintable
\  \ \|\
${y^3{\bar y}^3}$,  ${y^4{\bar y}^4}$, ${y^5{\bar y}^5}$,
${{y}^6{\bar y}^6}$
\ \|\ ${y^1{\bar y}^1}$,  ${y^2{\bar y}^2}$,
${\omega^5{\bar\omega}^5}$,
${{\omega}^6{\bar\omega}^6}$
\ \|\ ${\omega^1{\bar\omega}^1}$,  ${\omega^2{\bar\omega}^2}$,
${\omega^3{\bar\omega}^3}$,  ${{\omega}^4{\bar\omega}^4}$  \crthick
$\alpha$ \|
1, ~~~0, ~~~~0, ~~~~1 \|
0, ~~~0, ~~~~1, ~~~~0 \|
0, ~~~0, ~~~~0, ~~~~1 \nr
$\beta$ \|
0, ~~~0, ~~~~0, ~~~~1 \|
0, ~~~1, ~~~~1, ~~~~0 \|
1, ~~~0, ~~~~0, ~~~~0 \nr
$\gamma$ \|
1, ~~~1, ~~~~0, ~~~~0 \|\
1, ~~~0, ~~~~0, ~~~~0 \|
0, ~~~1, ~~~~0, ~~~~0  \endtable}
\hfill}
\smallskip
\parindent=0pt
\hangindent=39pt\hangafter=1
\baselineskip=18pt

{{\it Table 4.} A three generations ${SU(3)\times SU(2)\times U(1)^2}$
model, with $\vert\alpha_L(b_j)-\alpha_R(b_j)\vert=0~~,~~(j=1,2,3)$.
The electroweak doublets from the Neveu--Schwarz sector are projected out
and the color triplets remain in the physical spectrum. }

\vskip 2.5cm

\vfill
\eject
\input tables.tex
\nopagenumbers
\magnification=1000
\tolerance=1200



{\hfill
{\begintable
\  \ \|\ ${\psi^\mu}$ \ \|\ $\{{\chi^{12};\chi^{34};\chi^{56}}\}$ \ \|\
{${\bar\psi}^1$, ${\bar\psi}^2$, ${\bar\psi}^3$,
${\bar\psi}^4$, ${\bar\psi}^5$, ${\bar\eta}^1$,
${\bar\eta}^2$, ${\bar\eta}^3$} \ \|\
{${\bar\phi}^1$, ${\bar\phi}^2$, ${\bar\phi}^3$, ${\bar\phi}^4$,
${\bar\phi}^5$, ${\bar\phi}^6$, ${\bar\phi}^7$, ${\bar\phi}^8$} \crthick
${\alpha}$
\|\ 1 \| $\{1,~0,~0\}$ \|
1, ~~1, ~~1, ~~0, ~~0, ~~1, ~~0, ~~1 \|
1, ~~1, ~~1, ~~1, ~~0, ~~0, ~~0, ~~0 \nr
${\beta}$
\|\ 1 \|
$\{0,~1,~0\}$  \|
1, ~~1, ~~1, ~~0, ~~0, ~~0, ~~1, ~~1 \|
1, ~~1, ~~1, ~~1, ~~0, ~~0, ~~0, ~~0 \nr
${\gamma}$
\|\ 1 \| $\{0,~0,~1\}$  \|
{}~~$1\over2$, ~~$1\over2$, ~~$1\over2$, ~~$1\over2$,
{}~~$1\over2$, ~~$1\over2$, ~~$1\over2$, ~~$1\over2$ \|
 $1\over2$, ~~0, ~~1, ~~1,
{}~~$1\over2$,
{}~~$1\over2$, ~~$1\over2$, ~~0 \endtable}
\hfill}
\smallskip
{\hfill
{\begintable
\   \ \|\
${y^3{\bar y}^3}$,  ${y^4{\bar y}^4}$, ${y^5{\bar y}^5}$,
${{y}^6{\bar y}^6}$
\ \|\ ${y^1{\bar y}^1}$,  ${y^2{\bar y}^2}$,
${\omega^5{\bar\omega}^5}$,
${{\omega}^6{\bar\omega}^6}$
\ \|\ ${\omega^2{\omega}^3}$,  ${\omega^1{\bar\omega}^1}$,
${\omega^4{\bar\omega}^4}$,  ${{\bar\omega}^2{\bar\omega}^3}$ \crthick
${\alpha}$ \|
1, ~~~0, ~~~~0, ~~~~1 \|
0, ~~~0, ~~~~1, ~~~~0 \|
0, ~~~0, ~~~~1, ~~~~1 \nr
${\beta}$\|
0, ~~~0, ~~~~0, ~~~~1 \|\
0, ~~~1, ~~~~1, ~~~~0 \|
0, ~~~1, ~~~~0, ~~~~1 \nr
${\gamma}$\|
1, ~~~1, ~~~~0, ~~~~0 \|
1, ~~~1, ~~~~0, ~~~~0 \|
0, ~~~0, ~~~~0, ~~~~1 \endtable}
\hfill}
\smallskip

\parindent=0pt
\hangindent=39pt\hangafter=1
\baselineskip=18pt

{{\it Table 5.} A three generations
${SU(3)\times SU(2)\times U(1)^2}$ model.
The condition $\vert\alpha_L(b_j)-\alpha_R(b_j)\vert=1~~,$ is obeyed for
$j=3$ but not for $j=1,2$. Therefore, we obtain the color triplets $D_1$,
${\bar D}_1$, $D_2$, ${\bar D}_2$, and the electroweak doublets
$h_3$ and ${\bar h}_3$. The notation used is the notation of table 2.}
\vskip 1.5cm
\vfill
\eject

\bye